# Demonstration of an optical mixing technique to drive Kinetic Electrostatic Electron Nonlinear waves in laser produced plasmas

J. L. Kline,[1] B. B. Afeyan,[2] W. A. Bertsche,[3] N. A. Kurnit,[1] D. S. Montgomery,[1] R. P. Johnson,[1] and C. Niemann[4]

[1] *Los Alamos National Laboratory, Los Alamos, NM 87545*
[2] *Polymath Research Inc., Pleasanton, CA, 94566*
[3] *University of California at Berkeley, Berkeley, CA 94720*
[4] *Department of Electrical Engineering, University of California, Los Angeles, CA 90095*

A nitrogen gas Raman cell system has been constructed to shift a 70 J 527 nm laser beam to 600 nm with 20 J of energy. The 600 nm probe and a 200J, 527 nm pump beam were optically mixed in a laser produced (gas jet) plasma. The beating of the two laser beams formed a ponderomotive force that can drive Kinetic Electrostatic Electron Nonlinear (KEEN) waves discovered in Vlasov-Poisson simulations by Afeyan et al [1,2]. KEEN waves were detected in these experiments where traditional plasma theory would declare there to be a spectral gap (ie no linear waves possible). The detection was done using Thomson scattering with probe wavelengths of both 351 nm and 263.5 nm.

Since the first confirmed observation of nonlinear waves with frequencies significantly below that of electron plasma waves in a laser produced plasma [3], there has been great interest in the nature of these waves. Typically, in magnetic-field free, Maxwellian plasmas only two linear electrostatic waves exist, the electron plasma wave and the ion acoustic wave. However, when the electron velocity distribution function is modified, the dielectric plasma function changes making it possible for other waves to become weakly damped and grow. In laser produced plasmas, the incident laser light can resonantly couple with an electron plasma wave and a backscattered light wave, know as Stimulated Raman Scattering (SRS), driving the electron plasma wave to large amplitudes. Electrons can then be trapped in the electric potential of the electron plasmas wave accelerating them and thereby modifying the electron velocity distribution function [4]. To investigate the nature of this interaction, initial work invoked a BGK mode which involves the stationary limit of the Vlasov-Possion system of equations and relies on two populations of electrons, those trapped and un-trapped by the self-consistent wave potential [3]. While this approach finds that waves with frequencies slightly below that of linear electron plasma waves and also so called Electron Acoustic Waves (EAW) become sustainable, if a suitably modified velocity distribution function were posited, This does not address just how this modified distribution function arose and what can be done self-consistently as opposed to in an ad hoc manner. Vlasov-Poisson (and Vlasov-Maxwell) simulations starting from a Maxwellian have led to the discovery that in the spectral gap of traditional, textbook plasma physics, a new class of self-consistent waves, nonstationary Kinetic Electrostatic Electron Nonlinear (KEEN) waves [1,2] exist, and can be driven over a large region of ($\omega$,k) space between that of electron plasma waves and ion acoustics waves. In the case of laser plasma interactions, these nonlinear waves can interact with electron plasma waves driven by SRS and efficiently saturate the growth of the three-wave interaction, an important process for high energy density physics involving lasers, especially for inertial confinement fusion [5, 6]

To investigate KEEN waves in a controlled environment in which the waves can be either coupled to or decouple from SRS driven by a pump laser, optical mixing of two laser beams is employed to drive electrostatic waves from the ponderomotive force resulting from the beating of the two light waves. The desired frequency and wave number for the driven wave are given by the phase matching conditions, such that $\omega_{KEEN} = \omega_{pump} - \omega_{probe}$ and $\mathbf{k}_{KEEN} = \mathbf{k}_{pump} - \mathbf{k}_{probe}$. To solve this problem, a set of nitrogen Raman cells in an oscillator + amplifier configuration was designed and built to shift a 527 nm laser beam to 600 nm. While Raman shifters are typically used for very low energy laser systems, here the technique has been applied to a high energy laser system. In a Raman cell, light is scattered off the vibrational modes of a gas in a pressurized vessel, creating light at higher and lower frequencies with multiple orders such that, $\omega_s = \omega_o \pm n\omega_{vib}$ where $\omega_s$ is the scattered light frequency, $\omega_o$ is the incident laser light, n is the order and $\omega_{vib}$ is the vibrational frequency of the gas. We chose nitrogen in order to operate at the desired frequency with a wavelength of 600 nm. As an additional benefit, the second Stokes line was at 697 nm which could be used to drive electron plasma waves by optical mixing and study them independently of any SRS generated ones.

In this Letter, we demonstrate the first implementation of a Nitrogen Raman cell system for optical mixing of two lasers to drive KEEN waves. The Raman system converts a 70 J 527 nm laser into a 20 J 600 nm beam. The KEEN waves are detected using Thomson scattering. The successful demonstration of this technique opens the door to studying new nonlinear kinetic structures in laser produced plasmas.

The Raman probe laser system for producing 600 nm light consists of two Raman cells, one used as an oscillator and the other as an amplifier, both using the 2330 $N_2$ vibrational Raman transition [7]. While details of the Raman system will be described elsewhere, a brief description is provided here. The system starts by picking off a small piece of a 70 J, 527 nm, 8" diameter beam using a prism and collimating it into a 3mm size beam. This beam then passes twice through a 2 meter long 2" diameter pressure cell, filled with ~40 PSI of nitrogen, to produce a 600 nm seed beam. A mirror coated for 600 nm reflects the converted light and passes through all other wavelengths. The 600 nm light is then expanded to ~ 2" in diameter and made collinear with the remaining 527 nm light, also reduced to ~2" in diameter. The two beams pass through a 8" diameter pressure vessel three times amplifying the 600 nm seed through scattering off the 527 nm light. Again, a 600 nm coated mirror is used to pick out the 600 nm light. The output is then transported to the target chamber. Measurements of the focal spot showed that the output beam had a nearly diffraction limited quality.

The optical mixing KEEN wave experiments were performed at the Trident laser facility with the layout as shown in figure 2 [8]. A single 527 nm beam both generates the plasma and acts as one of the optical mixing beams. The 527 nm pump beam has 0 – 200 Js of energy in a 1ns square pulse. The beam is focused ~800 μm above the exit of a gas jet target to form the plasma by ionizing and heating the gas exiting the nozzle. The 527 nm pump light is focused with *f/*6 lenses through a 2mm hexagonal Random Phase Plate (RPP) [9] producing a zero-to-zero spot size of ~600 μm in diameter. The 600 nm Raman shifted probe beam passes through a square random phase plate and is focused in the gas jet plasma using an f/6 achromatic lens producing a zero-to-zero spot size of ~150 microns. The beams are crossed at an angle of 153 degrees to obtain the desired wave number. This geometry and the wavelengths generated by the Raman cell fix the frequency and wave number of the ponderomotive drives at KEEN wave: $\omega_{keen}$ = 4.35173 x $10^{14}$ rad/s and $|k_{KEEN}|$ = 217785 rad/cm.

There are several key diagnostics used to detect and measure the attributes of the ponderomotively driven KEEN waves, as well as measuring the experimental plasma parameters (Figure 2). The backscattered light and self-Thomson scattering diagnostics are used to measure plasma parameters during the period in which KEEN waves are driven. Thomson scattering and a transmitted beam diagnostic are used to measure the KEEN wave properties. The Thomson scattering diagnostics used both a 263.5 and 351 nm probe laser with Thomson scattering angles of 55.3º and 78.6º respectively. The 263.5 nm setup used a 200 ps probe pulse while the 351 nm probe setup use a 1 ns probe pulse. It should be noted that the 351 nm probe was adopted due to the reduced laser energy obtained in the 263.5 nm setup. Base on the wavelength of the probe and the Thomson scattering angles of the two configurations, the scattered light from the KEEN waves were expected to be observed at 280.4 nm for the 263.5 nm probe and at 382.3 for the 351 nm probe.

Spectrally integrated measurement of the Thomson scattering signals for both setups as a function of time is shown in Figure 3. Figure 3a shows the measured spectrum for the 263.5 nm Thomson scattering setup and Figure 3b shows the measured spectrum using the 351nm setup. Both show scattered light at the wavelength predicted from the Thomson scattering setup. Without the beam from the Raman system, the signals were not present. Such measurements were made by varying in the plasma conditions, as well as by varying the pump and probe laser energies. The waves were detected in both He and $N_2/H_2$ plasmas. Figure 4 shows the regions of ($\omega$,k) space in which driven KEEN waves were detected during these experiments. This is in contrast to the EAW attributed observations in the Montgomery *et. al* experiments, as well as an electron plasma wave. The dashed lines outline the locations in which the present experimental configuration can be used to drive KEENs or EPWs.

The initial results from the crossed beam experiments demonstrate success in driving and detecting KEEN waves using a nitrogen Raman system to convert a 527 nm laser beam to 600 nm. With the ability to drive and detect KEEN waves the nonlinear physics of these waves can be studied, as well as their interaction with SRS. It should also be noted that the Raman system described here could be used with other gases or in different Raman orders (for example $2^{nd}$ Stokes light at 697 nm) to study the nonlinear plasma response over a wider range of the ($\omega$, k) plane.

These experiments would not have been possible without the technical expertise of R. Gibson, F. Archuleta, R. Gonzales, T. Hurry, D. Esquibel, S. Reid, and T. Ortiz of the Trident laser crew. This work is supported by the DOE Stockpile Stewardship Academic Alliance program Polymath Research Inc. Grant DE-FG03-03NA00059 and under the auspices of the DOE/NNSA by LANL under Contract No. W-7405-ENG-36.

**Captions**

**Figure 1**: Schematic layout of the experiment with diagnostics.

**Figure 2:** Schematic layout of the Raman system.

**Figure 3:** a) Thomson scattering streak measurement of a KEEN wave using 263.5 nm probe. b) Thomson scattering streak measurement of a KEEN wave using 351.0 nm probe.

**Figure 4:** Plot of normalize ($\omega$,k) space. The solid line represents the least-damped mode calculated using BGK theory. The solid colored regions marked with He and $N_2/H_2$ are the locations in which KEEN waves were driven during these experiments. The regions encircled by the dashed lines and marked 600 nm and 697 nm represent the area in $\omega$,k space believed to be possible with the current experimental setup using the nitrogen Raman cell and the first and second Stokes shifts.

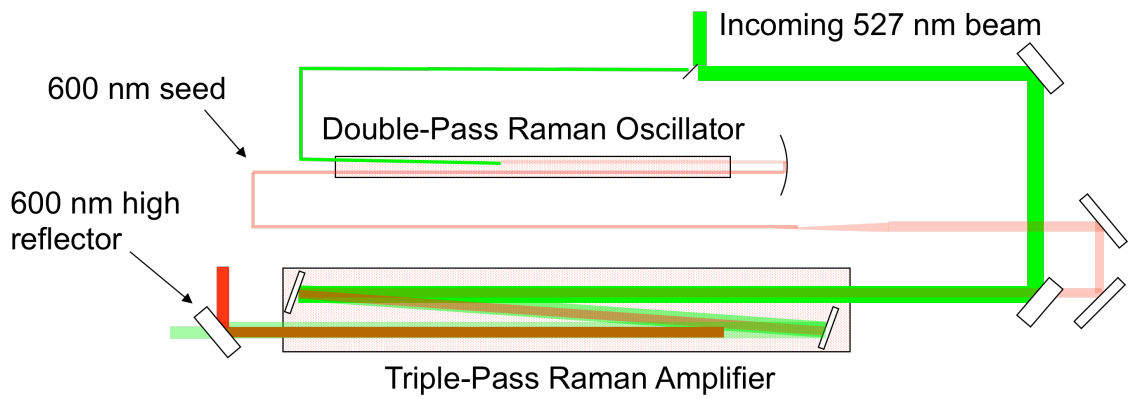

Kline figure 1

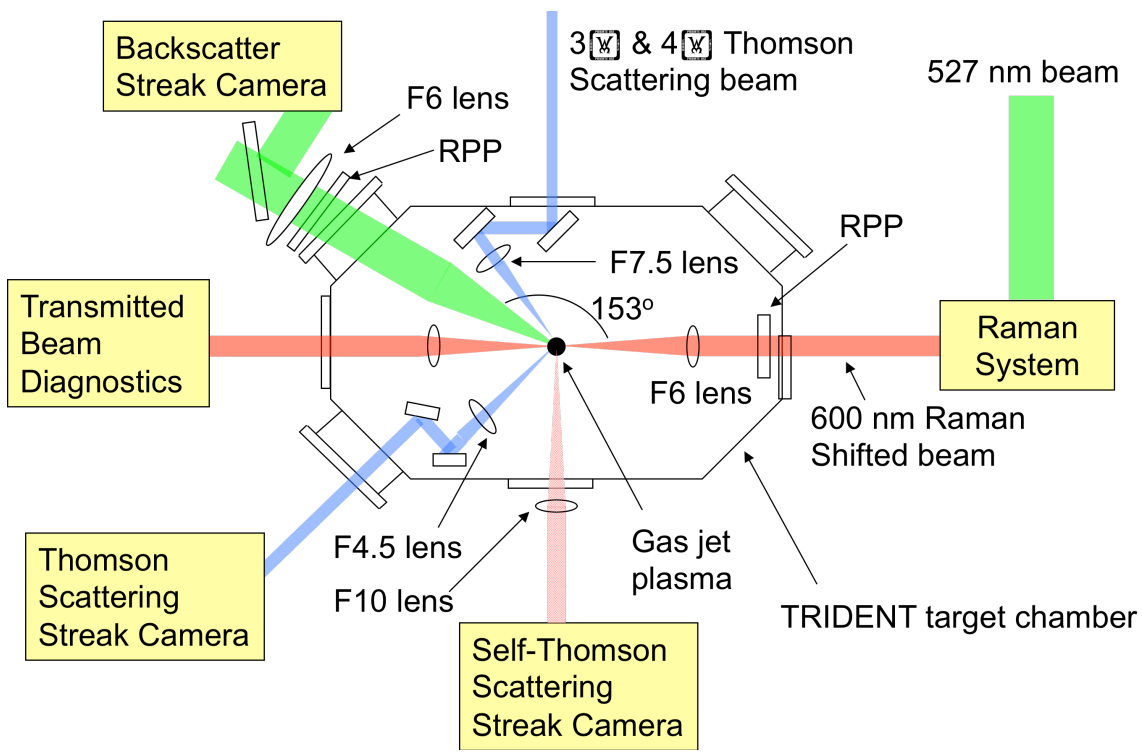

Kline figure 2

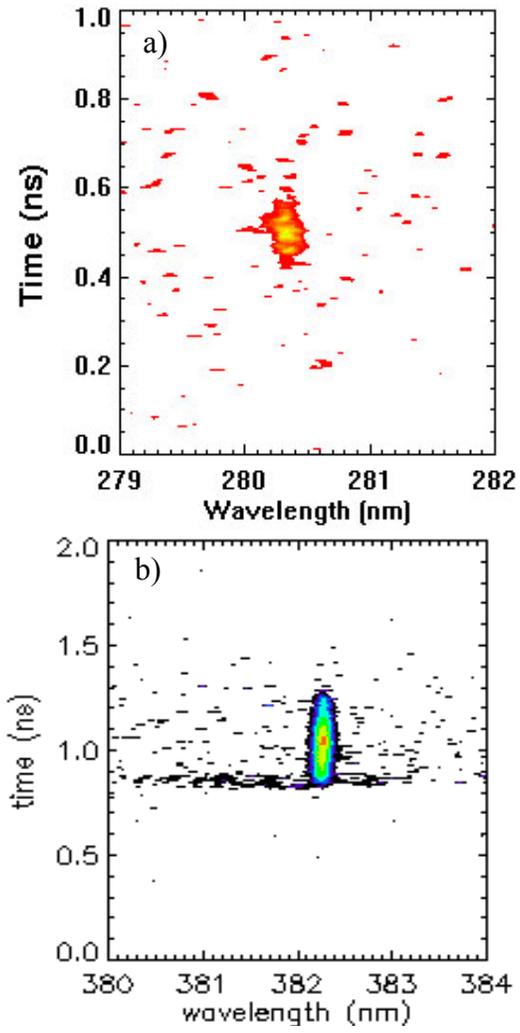

Kline Figure 3

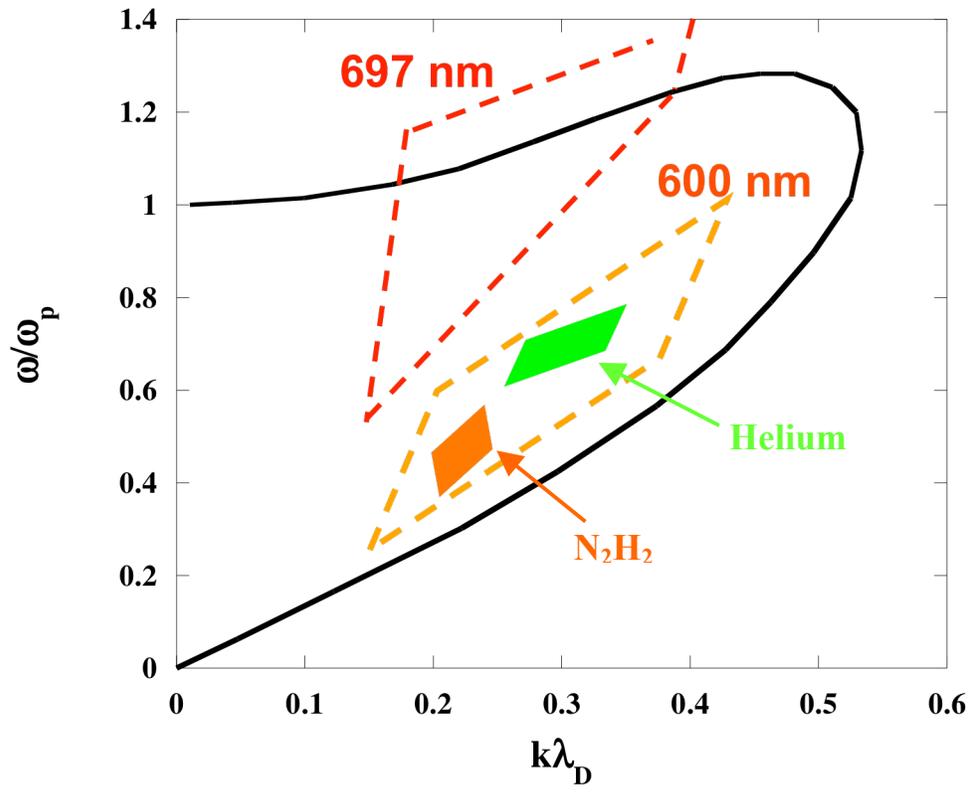

Kline figure 4